\documentclass[aPhyssubmission, Phys]{SciPost}

\DeclareUnicodeCharacter{03A6}{$\Phi$}

\hypersetup{
    colorlinks,
    linkcolor={red!50!black},
    citecolor={blue!50!black},
    urlcolor={blue!80!black}
}

\usepackage[bitstream-charter]{mathdesign}
\usepackage[capitalise]{cleveref}
\usepackage{graphicx}
\usepackage{subcaption}
\usepackage{float}
\DeclareSymbolFont{usualmathcal}{OMS}{cmsy}{m}{n}
\DeclareSymbolFontAlphabet{\mathcal}{usualmathcal}

\begin{document}

\begin{center}{\Large \textbf{
Dark sector studies with the PADME experiment\\
}}\end{center}

\begin{center}
A.P. Caricato\textsuperscript{1,2}, M. Martino\textsuperscript{1,2}, I. Oceano\textsuperscript{1,2},  F. Oliva\textsuperscript{1,2},  S. Spagnolo\textsuperscript{1,2}, G. Chiodini\textsuperscript{2}, F. Bossi\textsuperscript{3}, R. De Sangro\textsuperscript{3}, C. Di Giulio\textsuperscript{3}, D. Domenici\textsuperscript{3}, G. Finocchiaro\textsuperscript{3}, L.G. Foggetta\textsuperscript{3}, M. Garattini\textsuperscript{3}, A. Ghigo\textsuperscript{3}, F. Giacchino\textsuperscript{3}, P. Gianotti\textsuperscript{3}, T. Spadaro\textsuperscript{3}, E. Spiriti\textsuperscript{3}, C. Taruggi\textsuperscript{3}, E. Vilucchi\textsuperscript{3}, V. Kozhuharov\textsuperscript{3,4}, S. Ivanov\textsuperscript{4}, Sv. Ivanov\textsuperscript{4}, R. Simeonov\textsuperscript{4}, G. Georgiev\textsuperscript{4,5}, F. Ferrarotto\textsuperscript{6}, E. Leonardi\textsuperscript{6}, P. Valente\textsuperscript{6}, E. Long\textsuperscript{6,7$\star$}, G.C. Organtini\textsuperscript{6,7}, G. Piperno\textsuperscript{6,7}, M. Raggi\textsuperscript{6,7}, S. Fiore\textsuperscript{6,8}, P. Branchini\textsuperscript{9}, D. Tagnani\textsuperscript{9}, V. Capirossi\textsuperscript{10}, F. Pinna\textsuperscript{10}, A. Frankenthal\textsuperscript{11}

\end{center}

\begin{center}
{\bf 1} INFN Lecce, Lecce, Italy
\\
{\bf 2} Dip. di Matematica e Fisica, Universit\`a del Salento, Lecce, Italy
\\
{\bf 3} INFN Laboratori Nazionali di Frascati, Frascati, Italy
\\
{\bf 4} University of Sofia ``St. Kl. Ohridski'', Sofia, Bulgaria
\\
{\bf 5} INRNE, BAS, Sofia, Bulgaria
\\
{\bf 6} INFN Roma1, Rome, Italy
\\
{\bf 7} Dip. di Fisica, ``Sapienza'' Universit\`a di Roma, Rome, Italy
\\
{\bf 8} ENEA centro ricerche, Frascati, Italy
\\
{\bf 9} INFN Roma3, Rome, Italy
\\
{\bf 10} Politecnico di Torino, Turin, Italy
\\
{\bf 11} Dep. of Physics Princeton University, Princeton, USA
\\
* elizabeth.long@uniroma1.it
\end{center}

\begin{center}
\today
\end{center}


\definecolor{palegray}{gray}{0.95}
\begin{center}
\colorbox{palegray}{
  \begin{tabular}{rr}
  \begin{minipage}{0.1\textwidth}
    \includegraphics[width=30mm]{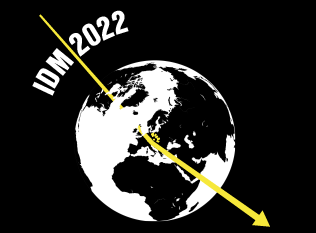}
  \end{minipage}
  &
  \begin{minipage}{0.85\textwidth}
    \begin{center}
    {\it 14th International Conference on Identification of Dark Matter}\\
    {\it Vienna, Austria, 18-22 July 2022} \\
    \doi{10.21468/SciPostPhysProc.?}\\
    \end{center}
  \end{minipage}
\end{tabular}
}
\end{center}

\section*{Abstract}
{\bf
The Positron Annihilation to Dark Matter Experiment (PADME) uses the positron beam of the DA$\Phi$NE Beam-Test Facility, at the Laboratori Nazionali di Frascati (LNF) to search for a Dark Photon $A'$. The search technique studies the missing mass spectrum of single-photon final states in $e^+e^-\rightarrow A'\gamma$ annihilation in a positron-on-thin-target experiment.
This approach facilitates searches for new particles such as long lived Axion-Like-Particles, protophobic X bosons and Dark Higgs.
This talk illustrated the scientific program of the experiment and its first physics results. In particular, the measurement of the cross-section of the SM process $e^+e^-\rightarrow \gamma\gamma$ at $\sqrt{s}$=21 MeV was shown.}

\vspace{10pt}
\noindent\rule{\textwidth}{1pt}
\vspace{10pt}

\section{Introduction}
\label{sec:intro}
\begin{figure}[H]
	\begin{subfigure}{0.43\linewidth}
		\includegraphics[scale=0.57]{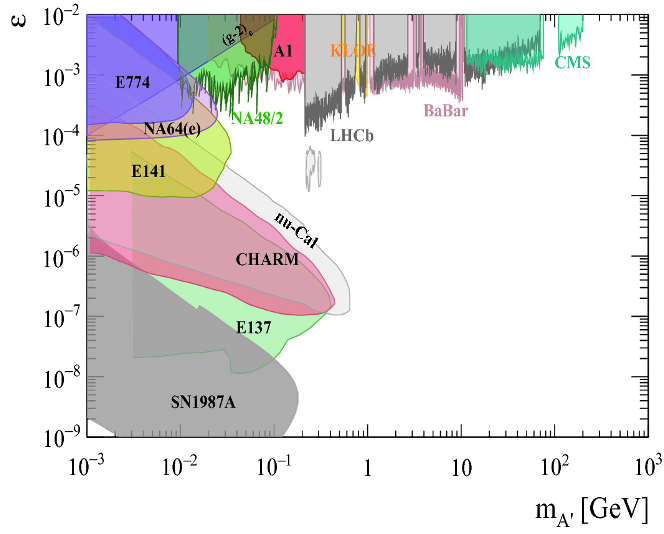} 
       \caption{Constraints for visibly decaying $A'$s.}
       \label{subfig:VisibleSearches}
        \centering
	\end{subfigure}
	\hfill
	\begin{subfigure}{0.43\linewidth}
		\includegraphics[scale=0.57]{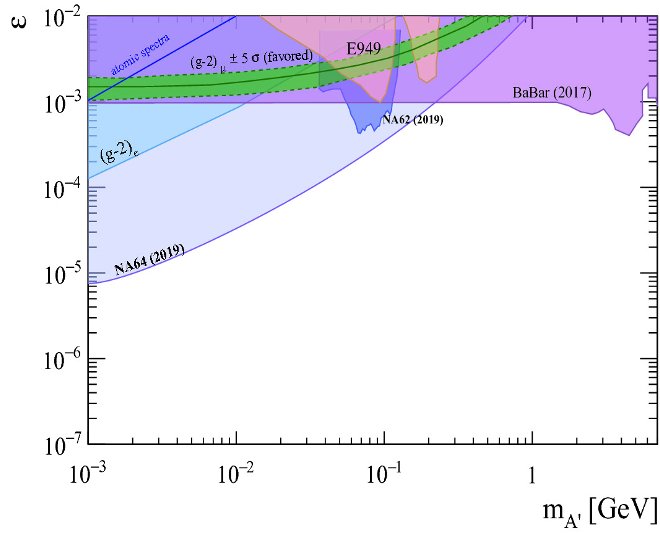} 
       \caption{Constraints for invisibly decaying $A'$s.}
       \label{subfig:InvisibleSearches}
        \centering
	\end{subfigure}
	\caption{Current constraints on Dark Photon models \cite{Fabbrichesi:2020wbt}. Models are characterised by two variables: the mass, $m_{A'}$, and the coupling, $\varepsilon$, between the $A'$ and SM leptons.}
	\label{fig:DPConstraints}
\end{figure}

With the restricted parameter space available to Weakly Interacting Massive Particle (WIMP) models, attention has turned to models in which Dark Matter is made of feebly interacting particles with masses below the electroweak symmetry-breaking scale. Several extensions of the Standard Model (SM) include a new Dark Sector governed by a $U_1(D)$ symmetry, in which feebly interacting Dark Matter particles interact with SM particles exclusively through the exchange of a particle, known as a ``Dark Photon'' ($A'$). These models are characterised by two parameters: the mass of the $A'$ ($m_{A'}$) and its coupling ($\varepsilon$) to Standard Model particles. Current constraints for Dark Photon models can be seen in \Cref{subfig:VisibleSearches,subfig:InvisibleSearches} respectively \cite{Fabbrichesi:2020wbt}.

\section{The PADME Experiment}\label{Sec:PADMEExperiment}

The positron beam from the Beam Test Facility at the National Laboratories of Frascati (LNF) impinges with a maximum energy of 550~MeV on the Active Diamond Target of the Positron Annihiltaion to Dark Matter Experiment (PADME). The 100~$\mu$m thick Target is used to measure beam intensity and position.

\begin{figure}[H]
	\captionsetup{width=0.45\textwidth}
	\begin{minipage}[t]{0.45\textwidth}
		\includegraphics[scale=0.2]{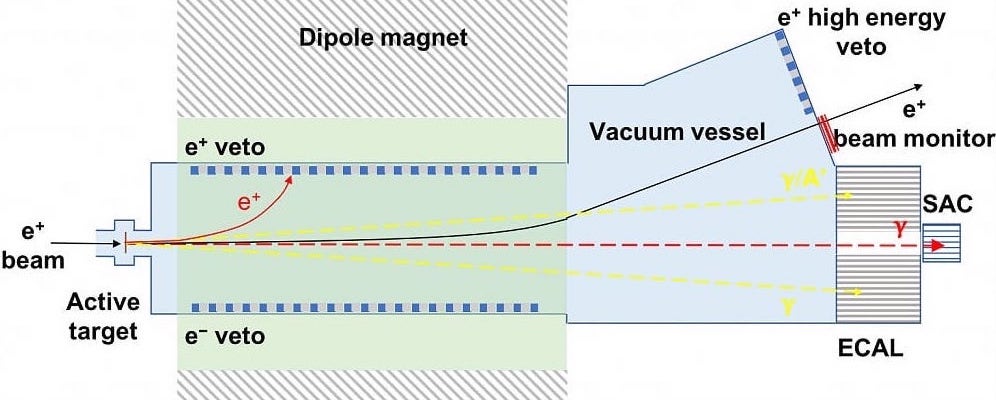}
 	    \caption{Schematic of PADME setup.}
    		\label{fig:PADMESchema}
    	\centering
	\end{minipage}
	\hfill 
	\begin{minipage}[t]{0.45\textwidth}
	\centering
    \includegraphics[scale=0.75]{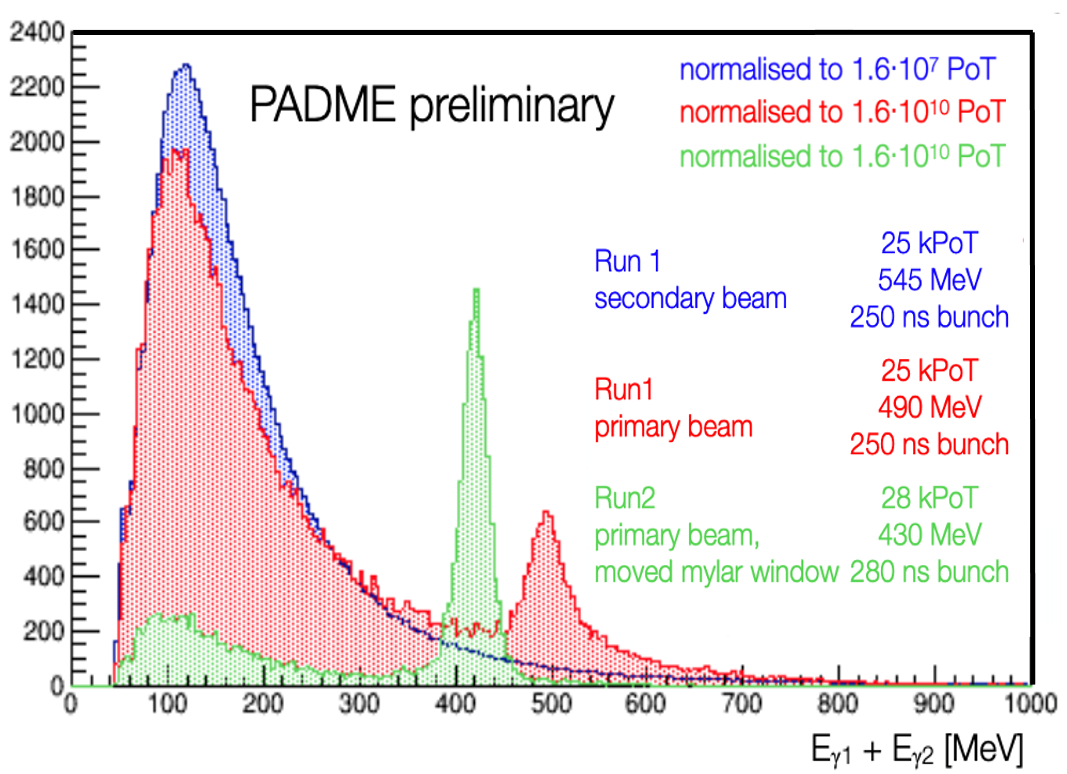} 
    \caption{Sum of energy of two photons in ECal. Run I primary (red) and secondary (blue) beam.  Run II primary beam and changed vaccuum separation (green). }
    \label{fig:TotECalEnergy}
	\end{minipage}
\end{figure}

In $e^+e^-$ collisions, three Dark Photon production mechanisms are available: resonant production, ``associated production'' in which one SM photon in two-photon annihilation is replaced with an $A'$, and ``$A'$-sstrahlung'' - analgous to Bremsstrahlung with the SM photon replaced by an $A'$. PADME was designed to search for associated production, the signal of which is a single SM photon in the BGO Electromagnetic Calorimeter (ECal) and nothing in the other detectors, as seen in \Cref{fig:PADMESchema}. The mass $m_{A'}$ is then calculated from the missing energy.

The main background is Bremsstrahlung. Since the Bremsstrahlung photon angle spectrum peaks sharply at low angles to the beam, the ECal was built with a central hole, behind which is the Small Angle Calorimeter (SAC), made from PbF$_2$, which has a faster response time than the ECal, enabling resolution of the high flux of Bremsstrahlung photons. The SAC is used with the Charged Particle Vetoes, made of plastic scintillator, two of which (the Positron Veto and Electron Veto) are located inside the magnetic field which directs charged final-state particles into the Vetoes, and un-interacted positrons into the TimePix3 beam monitor. More information about the detector can be found in \cite{PADMEComissioning}.

While PADME was designed to search for Dark Photons which decay invisibly, the Charged Particle Vetoes give access to visible final states where the $A'$ decays through $A'\rightarrow e^+e^-$.

\section{PADME Data Taking}
Between October 2018, when PADME was first switched on, and the end of Run II in 2020, more than $10^{13}$ positrons on target (PoT) were acquired in three beamline configurations. The total energy of photon pairs reaching the ECal can be seen in \Cref{fig:TotECalEnergy}. 

The first data in Run I was taken using the secondary positron beam of the LNF LINAC using positrons produced at a 1.7 X$_0$ Cu target close to the PADME hall. This resulted in a large number of low-energy particles entering the ECal. Moving to the primary beam, produced at the LINAC positron converter further upstream, a peak in the energy spectrum in the ECal is observed at the beam energy, however the long tails show significant pile-up and there was still a high background of low-energy particles.

Between Run I and Run II, the beryllium window separating the PADME vacuum from the LINAC vacuum was changed to Mylar and moved upstream of a wall which separates the LINAC and BTF. This produced significantly less low-energy beam-induced background and a much sharper peak at the beam energy.

More information about the two types of positron production and the comissioning of the vaccuum separation can be found in \cite{PADMEComissioning} and \cite{PADMEMC}.


\section{Multiple Photon Annihilation}
\sloppy
The first physics measurement at PADME is the inclusive in-flight cross section $\sigma(e^+e^-\rightarrow\gamma\gamma(\gamma))$, an important measurement for the community and for PADME for the following reasons: it could itself be sensitive to Sub-GeV scale new physics, for example in the form of Axion Like Particles (ALPs); there are no meausrements of this cross section with precision <20\% for beam energies below 500~MeV, and the measurements that do exist below this energy measure $e^+e^- \rightarrow non-charged\ particles$ in bubble chambers and would therefore be unable to distinguish SM contributions from new physics contributions; it allowed for full characterisation of the ECal; the cross section of associated production of a Dark Photon is linearly correlated to the cross section of 2$\gamma$ $e^+e^-$ annihilation ($\sigma(e^+e^-\rightarrow\gamma\gamma)$) as:
$$\sigma(e^+e^-\rightarrow\gamma A')\propto \varepsilon^2\times\sigma(e^+e^-\rightarrow\gamma\gamma)\times\delta(m_{A'})$$
where $\varepsilon$ is the coupling constant of the $A'$ to SM leptons and $\delta(m_{A'})$ is an acceptance factor that is a function of the mass of the Dark Photon.

The measurement was made with 10\% of the Run II data using a tag-and-probe method, exploiting the two-body kinematics of 2$\gamma$ annihilation, where the energy, $E$, of a photon is a function, $f(\theta)$, of its angle, $\theta$, from the beamline. Photons satisfying the condition $E_1-f(\theta_1)\sim 0$ were used as ``tags'' and matched with ``probes'' which had energy $E_1+E_2=E_{beam}$.

\sloppy 
The runs used had a number of Positrons on Target per bunch (PoT/bunch) stable across time between 19,000 and 36,000~PoT/bunch, and beam energy of 430~MeV. A total of $4\times10^{11}$~PoT were used in the final analysis. Selected events contained at least two good quality clusters in the ECal, where cluster quality was determined using the following criteria: Time difference between clusters |$\Delta T|<10$~ns; energy of each cluster $E_\gamma>$90~MeV; the difference between energy measured by the ECal and energy measured as a function of angle from the beam, $\Delta E(\theta) = E_\gamma - f(\theta)<$100~MeV; the leading photon had to fall in the fiducial region (115.8$< R<$285~mm) of the ECal.

The measurement of the inclusive cross-section $\sigma(e^+e^-\rightarrow\gamma\gamma(\gamma))$ at PADME is
$$\sigma(e^+e^-\rightarrow\gamma\gamma(\gamma)) = 1.977 \pm 0.018\ (stat) \pm 0.119\ (syst)\ \textrm{mb,}$$ while QED at next to leading order (NLO) is
$$\sigma(e^+e^-\rightarrow\gamma\gamma(\gamma)) = 1.9478 \pm 0.0005\ (stat) \pm 0.0020\ (syst)\ \textrm{mb,}$$ as given by BabaYaga, based on the calculations in \cite{BALOSSINI2008209}. Full details of the measurement are given in \cite{PADME:2022tqr}.

\begin{figure}[H]
    \centering
    \includegraphics[scale=0.25]{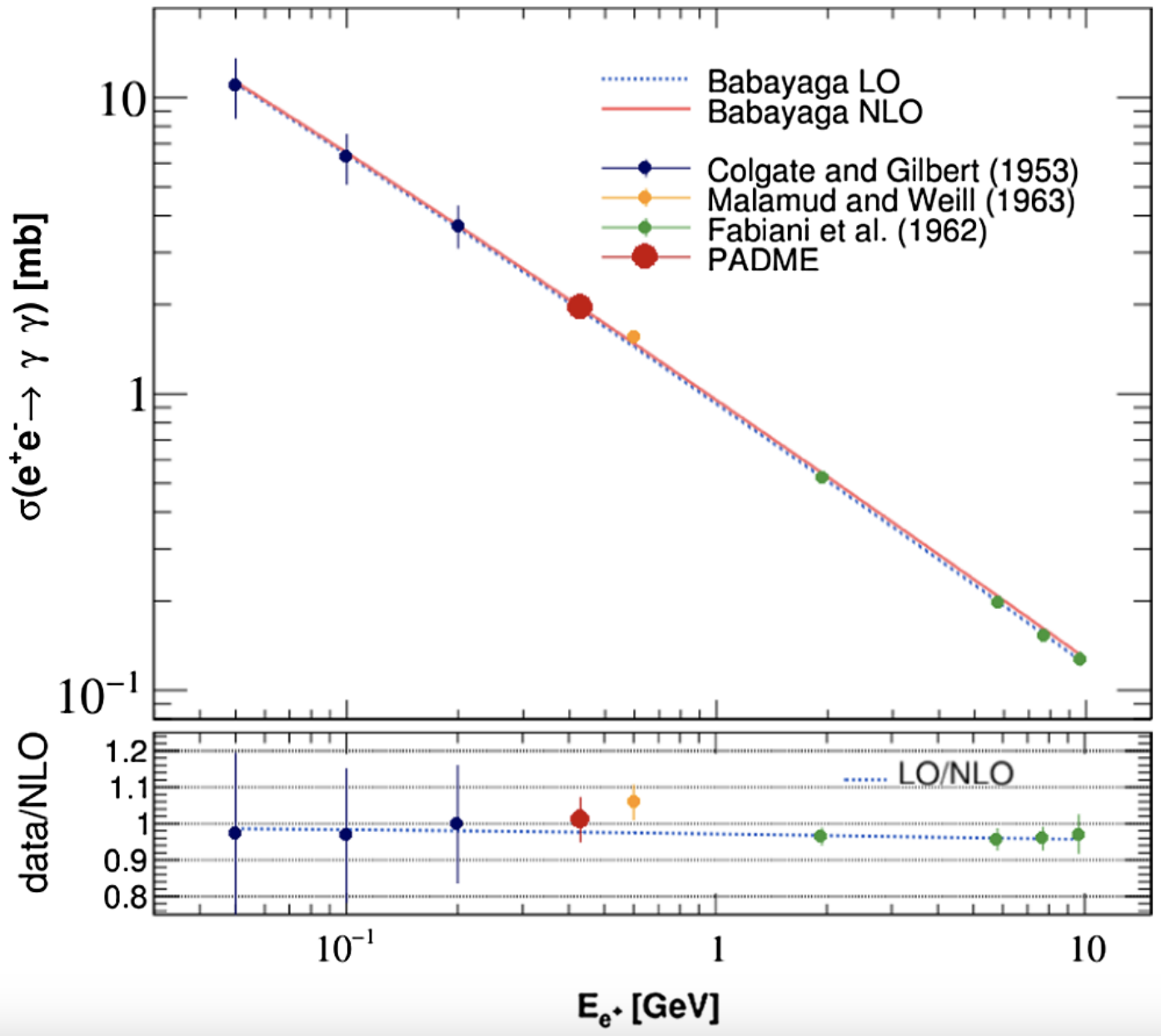} 
    \caption{Cross section of $\sigma(e^+e^-\rightarrow\gamma\gamma)$ \cite{PADME:2022tqr}.}
    \label{fig:CrossSectionMeasurement}
\end{figure}

The PADME measurement can be seen in context in \Cref{fig:CrossSectionMeasurement}.

\section{Padme Run III}
In 2016, A. J. Krasznahorkay et al.~announced the observation of a 6.8$\sigma$ anomaly in the opening angle between electrons and positrons from internal pair conversion (IPC) decays of excited $^8$Be nuclei \cite{PhysRevLett.116.042501}. In the experiments at the ATOMKI Institute in Debrecen, Hungary, $^7$Li nuclei were bombarded with protons to produce excited $^8$Be, which decays through IPC with a branching fraction of $\sim10^{-8}$.  The observed excess is compatible with a new particle (known as the ``X17") of mass 17~MeV decaying to $e^+e^-$ and was interpreted by Feng et al.~as an indication of a protophobic fifth fource of nature \cite{Feng:2016jff}. In 2019 an improved experimental setup was used to study IPC decays of three different excited states of $^4$He. These measurements confirmed the anomaly, with significances of 7.3$\sigma$, 6.6$\sigma$, and 8.9$\sigma$ respectively \cite{Krasznahorkay:2019lyl}.

\begin{figure}[H]
	\captionsetup{width=0.45\textwidth}
	\begin{minipage}[t]{0.45\textwidth}
		\includegraphics[scale=0.6]{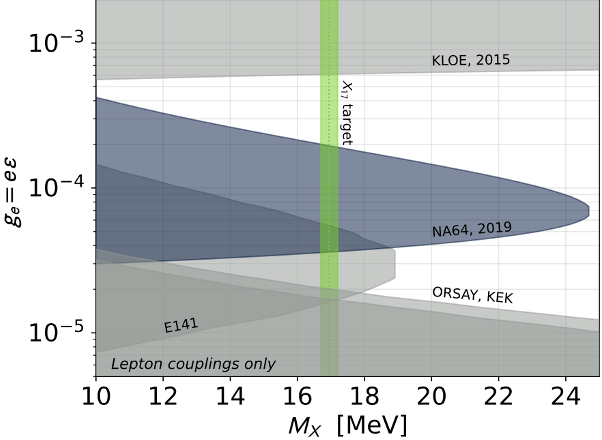} 
		\caption{Targetted parameter space for X17 particle.}
		\label{fig:X17Exclusion}
    	\centering
	\end{minipage}
	\hfill 
	\begin{minipage}[t]{0.45\textwidth}
		\includegraphics[scale=0.6]{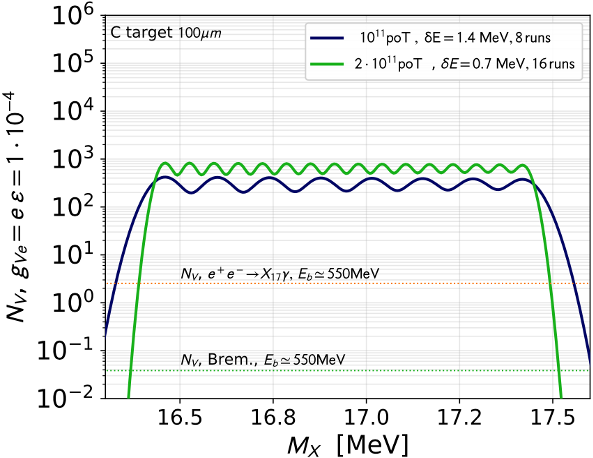} 
		\caption{Number of X17 expected at PADME at each point in energy scan.}		           		
		\label{fig:PADMEX17Scan}
	\end{minipage}
\end{figure}

These measurements' high sigificances and compatibility with each other led the PADME collaboration to search for the X17. The Beam Test Facility at LNF is the only facility in the world with a positron beam at low enough energy to produce this particle on resonance, increasing the production cross section by orders of magnitude. This means several thousand X17 particles would be produced in $10^{11}$ PoT even with small couplings of order $10^{-4}$ \cite{LucDarme}. PADME Run III will perform a scan over the mass region predicted by the ATOMKI experiments, shown in \Cref{fig:PADMEX17Scan}, looking for signatures of X17 particles decaying to $e^+e^-$ pairs.

In Run II data it was observed that the opening angle in S-channel Bhabha scattering made searching for $e^+e^-$ pairs in the Veto detectors very challenging. In Run III therefore, the magnetic field is switched off sending all particles towards the ECal. To distinguish between $e^+/e^-$ and photons, a new detector, the ``ETagger'', was built in front of the ECal using bars of 5~mm plastic scintillator. Run III started in July 2022 and will continue in the autumn.

\section{Conclusion}
PADME was built to search for Dark Photons with mass $\leq$ 23.7~MeV. In Runs I and II, more than $10^{13}$ Positrons on Target were collected, allowing the collaboration to perform the most precise measurement of the inclusive in-flight cross section $\sigma(e^+e^-\rightarrow\gamma\gamma(\gamma))$ for beam energies <500~MeV. The measurement is in agreement with the SM at NLO, as calculated by BabaYaga.

PADME Run III started in July 2022 and continues in the autumn in an attempt to confirm the existence of the X17 particle reportedly found in nuclear physics experiments at the ATOMKI insitute.



\paragraph{Funding information}
This work has been mainly funded by Istituto Nazionale di Fisica Nucleare. Other funds have been also granted by: the Italian Ministry of Foreign Affairs and International Cooperation (MAECI) under the grant PRG00226 (CUP I86D16000060005), the BG-NSF KP-06-DO02/4 from 15.12.2020 as part of MUCCA, CHIST-ERA-19-XAI-009, and TA-LNF as part of STRONG-2020 EU Grant Agreement 824093 projects.

\bibliography{2022IDMProceedings_BethLong.bib}

\begin{thebibliography}{1}
\providecommand{\url}[1]{\texttt{#1}}
\providecommand{\urlprefix}{URL }
\expandafter\ifx\csname urlstyle\endcsname\relax
  \providecommand{\doi}[1]{doi:\discretionary{}{}{}#1}\else
  \providecommand{\doi}{doi:\discretionary{}{}{}\begingroup
  \urlstyle{rm}\Url}\fi
\providecommand{\eprint}[2][]{\url{#2}}

\bibitem{Fabbrichesi:2020wbt}
M.~Fabbrichesi, E.~Gabrielli and G.~Lanfranchi,
\newblock \emph{{The Dark Photon}}  (2020),
\newblock \doi{10.1007/978-3-030-62519-1},
\newblock \eprint{2005.01515}.

\bibitem{PADMEComissioning}
P.~Albicocco, R.~Assiro, F.~Bossi, P.~Branchini, B.~Buonomo, V.~Capirossi,
  E.~Capitolo, C.~Capoccia, A.~P. Caricato, S.~Ceravolo, G.~Chiodini,
  G.~Corradi \emph{et~al.},
\newblock \emph{Commissioning of the padme experiment with a positron beam},
\newblock \doi{10.48550/ARXIV.2205.03430} (2022).

\bibitem{PADMEMC}
F.~Bossi, P.~Branchini, B.~Buonomo, V.~Capirossi, A.~Caricato, G.~Chiodini,
  R.~de~Sangro, C.~Di~Giulio, D.~Domenici, F.~Ferrarotto, S.~Fiore,
  G.~Finocchiaro \emph{et~al.},
\newblock \emph{The padme beam line monte carlo simulation}  (2022).

\bibitem{BALOSSINI2008209}
G.~Balossini, C.~Bignamini, C.~{Carloni Calame}, G.~Montagna, O.~Nicrosini and
  F.~Piccinini,
\newblock \emph{Photon pair production at flavour factories with per mille
  accuracy},
\newblock Physics Letters B \textbf{663}(3), 209 (2008),
\newblock \doi{https://doi.org/10.1016/j.physletb.2008.04.007}.

\bibitem{PADME:2022tqr}
F.~Bossi \emph{et~al.},
\newblock \emph{{Cross-section measurement of two-photon in-flight annihilation
  of positrons at s=20\,\,MeV with the PADME detector}},
\newblock Phys. Rev. D \textbf{107}(1), 012008 (2023),
\newblock \doi{10.1103/PhysRevD.107.012008},
\newblock \eprint{2210.14603}.

\bibitem{PhysRevLett.116.042501}
A.~J. Krasznahorkay, M.~Csatl\'os, L.~Csige, Z.~G\'acsi, J.~Guly\'as,
  M.~Hunyadi, I.~Kuti, B.~M. Nyak\'o, L.~Stuhl, J.~Tim\'ar, T.~G. Tornyi,
  Z.~Vajta \emph{et~al.},
\newblock \emph{Observation of anomalous internal pair creation in
  $^{8}\mathrm{Be}$: A possible indication of a light, neutral boson},
\newblock Phys. Rev. Lett. \textbf{116}, 042501 (2016),
\newblock \doi{10.1103/PhysRevLett.116.042501}.

\bibitem{Feng:2016jff}
J.~L. Feng, B.~Fornal, I.~Galon, S.~Gardner, J.~Smolinsky, T.~M.~P. Tait and
  P.~Tanedo,
\newblock \emph{{Protophobic Fifth-Force Interpretation of the Observed Anomaly
  in $^8$Be Nuclear Transitions}},
\newblock Phys. Rev. Lett. \textbf{117}(7), 071803 (2016),
\newblock \doi{10.1103/PhysRevLett.117.071803},
\newblock \eprint{1604.07411}.

\bibitem{Krasznahorkay:2019lyl}
A.~J. Krasznahorkay \emph{et~al.},
\newblock \emph{{New evidence supporting the existence of the hypothetic X17
  particle}}  (2019),
\newblock \eprint{1910.10459}.

\bibitem{LucDarme}
L.~Darmé,
\newblock \emph{X17 production mechanisms at accelerators}.

\end{thebibliography}

\nolinenumbers

\end{document}